\documentclass[review]{elsarticle}
\usepackage{subfigure}
\usepackage{lineno,hyperref}
\usepackage{indentfirst}
\usepackage{bm}
\usepackage{threeparttable}
\usepackage{multirow}
\usepackage{amsmath}
\usepackage{mathrsfs}
\usepackage{array}
\usepackage{amssymb}
\usepackage{graphicx}
\usepackage{url}
\usepackage{hyperref}
\usepackage{epstopdf}
\usepackage{booktabs}
\usepackage{color}
\usepackage{lscape}
\usepackage{geometry}
\usepackage{fancyhdr}

\journal{???}

\begin{document}

\begin{frontmatter}

\title{A direct unified wave-particle method for simulating non-equilibrium flows}
\author{Junzhe Cao$^a$}
\ead{caojunzhe@mail.nwpu.edu.cn}

\author[]{Sha Liu$^a$$^b$\corref{mycorrespondingauthor}}
\cortext[mycorrespondingauthor]{Corresponding author}
\ead{shaliu.@nwpu.edu.cn}

\author{Sirui Yang$^a$}
\ead{caojunzhe@mail.nwpu.edu.cn}

\author[]{Chengwen Zhong$^a$$^b$}
\ead{zhongcw@nwpu.edu.cn}


\address{$^a$School of Aeronautics, Northwestern Polytechnical University, Xi'an, Shaanxi 710072, China\\
$^b$Institute of Extreme Mechanics, Northwestern Polytechnical University, Xi'an, Shaanxi 710072, China}

\begin{abstract}
In this work, the Navier-Stokes (NS) solver is combined with the Direct simulation Monte Carlo (DSMC) solver in a direct way, under the wave-particle formulation [J. Comput. Phys. 401, 108977 (2020)]. Different from the classical domain decomposition method with buffer zone for overlap, in the proposed direct unified wave-particle (DUWP) method, the NS solver is coupled with DSMC solver on the level of algorithm. Automatically, in the rarefied flow regime, the DSMC solver leads the simulation, while the NS solver leads the continuum flow simulation. Thus advantages of accuracy and efficiency are both taken. At internal flow regimes, like the transition flow regime, the method is accurate as well because a kind of mesoscopic modeling is proposed in this work, which gives the DUWP method the multi-scale property. Specifically, as to the collision process, at $t < \tau$, it is supposed that only single collision happens, and the collision term of DSMC is just used. At $t > \tau$, it is derived that $1-\tau/\Delta t$ of particles should experience multiple collisions, which will be absorbed into the wave part and calculated by the NS solver. Then the DSMC and NS solver can be coupled in a direct and simple way, bringing about multi-scale property. The governing equation is derived and named as multi-scale Boltzmann equation. Different from the original wave-particle method, in the proposed DUWP method, the wave-particle formulation is no more restricted by the Boltzmann-BGK type model and the enormous research findings of DSMC and NS solvers can be utilized into much more complicated flows, like the thermochemical non-equilibrium flow. In this work, one-dimensional cases in monatomic argon gas are preliminarily tested, such as shock structures and Sod shock tubes.
\end{abstract}


\end{frontmatter}

\section{Introduction}\label{Sec:introduction}
With the development of hypersonic vehicle, spacecraft and micro-electromechanical system, the multi-scale non-equilibrium flow receives more attention and it is a challenging topic to develop corresponding multi-scale numerical methods, which can accurately simulate flows from the continuum flow regime to the free molecular flow regime. In recent years, both the stochastic particle method \cite{pbgkre,dsmcpbgkre} and discrete velocity method (DVM) \cite{dvmre,imexre,ugksre,dugksre,idvmre,gsisre} are successfully developed. Nevertheless, one of the most important problems faced in these numerical methods is the efficiency problem. Though a lot of methods are proposed for acceleration or reducing memory consumption \cite{lvdsmc,apdsmc,csz1,csz2,yrf}, there is still need for further improvement. The root cause of efficiency problem is that when the non-equilibrium feature becomes stronger, there are rather more freedom degrees to be described. For example, the dimensionality curse of DVM makes the discrete velocity space quiet expensive for simulating hypersonic flows.

When the flow is rarefied enough, the Monte Carlo method is very efficiency to describe the velocity distribution function, which makes stochastic particle methods suitable for simulating rarefied flows, and the most famous method of them is the Direct simulation Monte Carlo (DSMC) method\cite{dsmcre}. It has been developed to maturity and used to simulate complex flows like thermochemical non-equilibrium flows in flow mechanism studies\cite{jfm1,jfm2} and engineering applications\cite{sparta,dsmcfoam}. However, the efficiency of stochastic particle methods reduces rapidly when the flow becomes near-continuum or low-speed. Generally, the number of particles to be numerically simulated should be prohibitively large when using stochastic particle methods to simulate near-continuum flows and low-speed flows, and the time step in DSMC is seriously restricted in the near-continuum flow regime. On the other hand, macroscopic methods for simulating continuum flows are adequately developed as well, based on Euler equations and Navier-Stokes (NS) equations\cite{jiri,toro,leveque}. These methods are efficient and accurate in the continuum flow regime, but invalid on rarefied flows. As a result, domain decomposition methods are developed to implement the two categories methods (macroscopic methods and stochastic particle methods, mainly the DSMC method) in their corresponding suitable domains, and set a buffet zone for overlap\cite{domaindecomp1,domaindecomp2,domaindecomp3,domaindecomp4,domaindecomp5}. These methods achieve success. However, the accuracy is not sufficient in transitional flows and buffer zones of unsteady multi-scale flows.

Recently, a unified gas-kinetic wave-particle (UGKWP) method is proposed \cite{ugkwp1,ugkwp2} on the framework of unified gas-kinetic scheme (UGKS)\cite{ugks1}. It is a multi-scale multi-efficiency preserving (MMP) method, and is rapidly developed \cite{ugkwp5,suwp,suwp2}. Benefitting from the analytical solution of Boltzmann-BGK type model, the stochastic particles are coupled with macroscopic methods (wave part, like the gas-kinetic scheme (GKS)\cite{gks}) in the level of algorithm, which is different from the empirically artificial divide of different methods on the mesh in domain decomposition methods. Because the UGKWP method is designed fully in accordance with the Boltzmann-BGK type model, it is accurate enough in all flow regimes, naturally better than domain decomposition methods, especially when simulating complex multi-scale flows, so the UGKWP achieves success on the study of complex multi-scale flow mechanism\cite{ugkwpa,ugkwpb}. However, aiming at more complicated multi-physical field multi-scale flows, for example, thermochemical non-equilibrium flows, it is a hard task to extend the Boltzmann-BGK type model. As to monatomic gas mixtures, the key of modeling is the recovery of three coefficients, affecting the diffusion, viscosity and heat flux. In the Andries-Aoki-Perthame (AAP) model\cite{gm3,gm4} and Ellipsoidal Statistical-BGK(ES-BGK) type model\cite{gm5,gm6}, there are a maximum of two correct transport coefficients. In the model proposed in Ref.\cite{gm8,gm9}, the three transport coefficients are recovered, but it is to be improved by incorporating the correct Soret and Dufour coefficients. As to polyatomic gas mixtures\cite{gm11,gm12} or even reactive mixtures\cite{gm13,gm14,gm15}, much more relaxation rates of macroscopic variable need to be recovered, like the relaxation rates of internal energy (rotational and vibrational energy). Some work has been done to deal with these problems. In Ref.\cite{dr}, a direct relaxation (DR) process is proposed to decrease the difficulty in building the Boltzmann-BGK type model. The only parameters used in the DR are relaxation rates of macroscopic variable. In Ref.\cite{suwp}, the multi-scale mechanism of UGKWP method is simplified into the quantified model-competition (QMC) mechanism. Though the Boltzmann-BGK type model or DR is still needed, the algorithm is much simplified with acceptable loss of accuracy. In Ref.\cite{xiaocong}, instead of the hard work on improving the Boltzmann-BGK type model, the relaxation time $\tau$ of each particle is modified. And the early-rising problem of shock wave simulation is solved. The direct modeling theory\cite{dm} implied in it is desirable to be developed.

When the hypersonic vehicle or spacecraft flies at high Mach number and high attitude, the rarefication effect is strongly associated with thermochemical non-equilibrium effect, and it is of vital importance to develop a physical model or numerical method which can simulate this kind of flow efficiently and accurately. The wave-particle formulation is perfect but existing methods are all restricted by the Boltzmann-BGK type model. In this work, the stochastic unified method for multi-scale flows is reflected, and a methods called direct unified wave-particle (DUWP) method is proposed to get rid of the restriction. The procedure of DUWP method can be listed as: particle transfer part, macroscopic flux transfer part, particle collision part, and particle absorption or separation part. In the particle transfer part, particles are freely transported without collision. The macroscopic flux is directly calculated by NS solvers. In the particle collision part, the collision term of DSMC is just used. And in the particle absorption or separation part, a simple criteria is set for sampling. As a consequence, the DSMC and NS solver are coupled in a direct and simple way, bringing about multi-scale property. And the enormous research findings of DSMC and CFD can be utilized.

The remainder of this paper is organized as follows: the basis of gas kinetic theory is introduced in Sec.\ref{sec:gkt}. Sec.\ref{sec:solver} is the construction of DUWP method. The numerical test cases are conducted in Sec.\ref{sec:cases}. The conclusions are in Sec.\ref{sec:conclusion}.

\section{The gas kinetic theory}\label{sec:gkt}
In the gas-kinetic theory, the physical system is described by the distribution function $f\left(\bm{x},\bm{\xi},\bm{\eta},t\right)$ which means the probability density of particles that arrive at position $\mathbf{x}$ at time $t$ with velocity $\bm{\xi}$ and internal energy represented by the equivalent velocity $\bm{\eta}$. As to the equilibrium state, the distribution function is totally determined by macroscopic values, known as Maxwellian distribution function:
\begin{equation}\label{eq:maxwellian}
f = g = {\left( {\frac{1}{{2\pi RT}}} \right)^{\frac{{D + 3}}{2}}}\exp \left( { - \frac{{{\mathbf{c}} \cdot {\mathbf{c}} + {\bm{\eta }} \cdot {\bm{\eta }}}}{{2RT}}} \right),
\end{equation}
where $R$ is specific gas constant, $T$ is temperature, $D$ is the degree of freedom about the rotational and the vibrational motion of molecules, $\mathbf{c}=\bm{\xi}-\mathbf{U}$ is the peculiar velocity (the macroscopic velocity $\mathbf{U}$ is also known as the mean velocity of all particles). The macroscopic conservative variables $\bm{\Psi}=\left(\rho, \rho \mathbf{U}, \rho E\right)$(density, momentum and energy) can be directly obtained from the distribution function $f$ by the following relation: ($\bm{\varphi}=\left({ 1,\bm{\xi}, {1\over 2}\left({ \bm{\xi}\cdot\bm{\xi}+\bm{\eta}\cdot\bm{\eta} }\right)}\right)$)
\begin{equation}\label{eq:gkt_moment0}
\bm{\Psi} = \int_{R^3} \rho\bm{\varphi}f d\bm{\xi} \int_{R^D} d\bm{\eta}.
\end{equation}
The integral is called the ``moment'' in the gas-kinetic theory. The stress $\mathbf{S}$ and heat flux $\mathbf{Q}$ can also be obtained by the following moments:
\begin{equation}\label{eq:gkt_moment1}
\begin{aligned}
& {\mathbf{S}} = \int_{R^3} \rho\mathbf{cc}(f-g) d\bm{\xi} \int_{R^D} d\bm{\eta},\\
& {\mathbf{Q}} = \int_{R^3} \rho\mathbf{c}\left[{ {1\over 2}m\left({ \mathbf{c}\cdot\mathbf{c} + \bm{\eta}\cdot\bm{\eta} }\right)f }\right] d\bm{\xi} \int_{R^D} d\bm{\eta}.
\end{aligned}
\end{equation}

\section{A direct unified wave-particle method}\label{sec:solver}

\subsection{Modeling of collision source term}
\begin{figure}
	\centering
	\subfigure{
			\includegraphics[width=0.75 \textwidth]{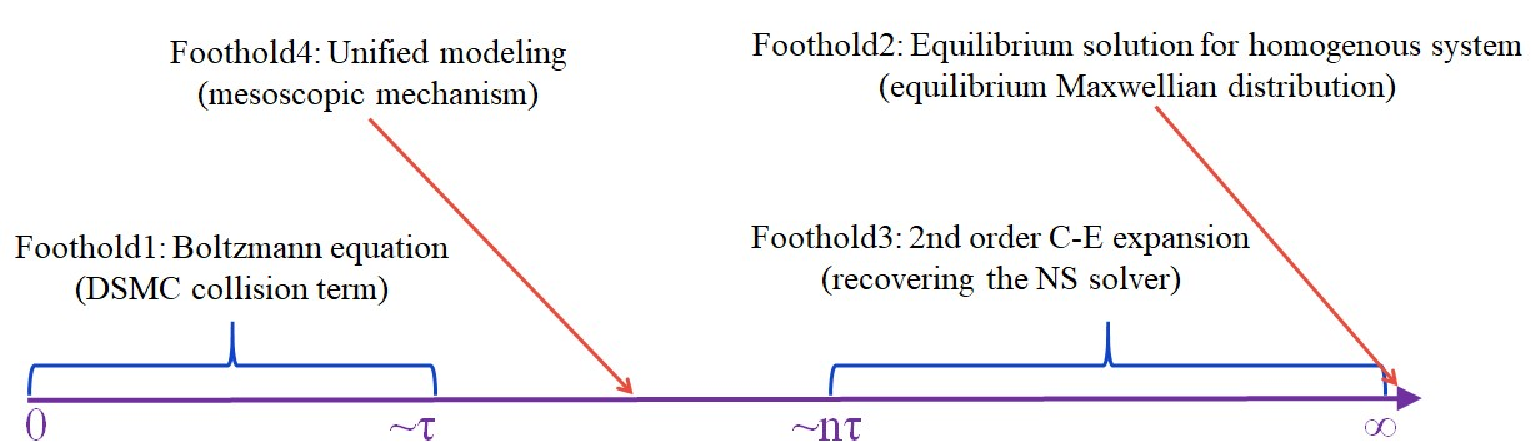}
		}
	\caption{\label{fig1} Modeling of collision source term.}
\end{figure}

The collision source term is described as Fig.\ref{fig1}. The coordinate is the evolution of time, divided into three parts, and $\tau=\frac{\mu}{p}$ is the molecular mean collision time ($\mu$ is the viscosity coefficient and $p$ is pressure). At $\left[{0,\tau}\right]$, it is supposed that only single collision happens. Microscopic mechanism leads this time interval and the DSMC method is used. At $\left[{n\tau, \infty}\right]$ ($n$ is large enough), the macroscopic mechanism leads the flow and the NS solver is used. At $\left[{\tau, n\tau}\right]$, it is supposed that multiple collision happens, and the mesoscopic model should be conducted.

The BGK model equation is taken for qualitative guide. Separating the transport term out, the BGK model equation becomes:
\begin{equation}
\frac{\partial f}{\partial t}=\frac{g-f}{\tau}.
\end{equation}
It is generally known that the Prantl number of BGK model is fixed to $1$, which means its relaxation rate of heat flux is always wrong, and the Shakhov-BGK model is a popular improved model, as:
\begin{equation}
\begin{aligned}
&\frac{\partial f}{\partial t}=\frac{g^S-f}{\tau},\\
&g^S=g\left[ {1+(1-\rm{Pr})\frac{\mathbf{c}\cdot\mathbf{Q}}{5pRT}\left( {\frac{|c|^2+|\eta|^2}{RT}-5} \right)} \right].
\end{aligned}
\end{equation}
However, as discussed in the introduction, as to thermochemical non-equilibrium flows, much more relaxation rates should be matched, thus the model will be prohibitively complicated or even unknown. In the DUWP method, the model is written as:
\begin{equation}
\frac{\partial f}{\partial t}=\frac{\mathscr{G}-f}{\tau}.
\end{equation}
It is assumed that $\mathscr{G}$ exists but it is not needed to get its specific expression, because the NS solver will be directly used to calculate this part. Its implicit differential solution is:
\begin{equation}\label{eq:ds}
f(t)=\frac{\tau}{\tau+t}f(0)+\frac{t}{\tau+t}\mathscr{G}.
\end{equation}
From Eq.\ref{eq:ds}, it can be derived that at $\left[{\tau, \Delta t}\right]$, $P$ of particles should be absorbed into the wave part ($\mathscr{G}$). $\Delta t$ is the time step. The expression of $P$ is:
\begin{equation}\label{eq:p}
P=
\left\{ \begin{array}{lr}
0, &\Delta t \leq \tau, \\
\frac{t}{\tau+t}|_{t=\Delta t-\tau}=\frac{\Delta t-\tau}{\Delta t-\tau+\tau}=1-\frac{\tau}{\Delta t}, &\Delta t > \tau.
\end{array}
\right.
\end{equation}
So finally the model is written as:
\begin{equation}\label{eq:model}
f(\Delta t)=P\mathscr{G}+\left({1-P}\right)B(f,f)\Delta t,
\end{equation}
where $B(f,f)$ denotes the collision term of Boltzmann equation (BE), which is calculated by DSMC.

\subsection{Multi-scale Boltzmann equation}
From another viewpoint, form in Eq.\ref{eq:model} can be derived from the BE,
\begin{equation}\label{eq:s1b1}
\frac{Df}{Dt}=B(f,f).
\end{equation}
On the other hand, although as a model for BE, a relaxation model equation such as Boltzmann-BGK equation is only accurate in continuum and near continuum regime, but its temporal integral solution has a concrete multi-scale property and clear physical meaning that the distribution of molecules tends to the local equilibrium along their trajectory, being consistent with the second law of thermodynamics. It is written as:
\begin{equation}
f\left({ t+\Delta t }\right)=e^{-\Delta t/\tau}f\left(t\right)+\left({ 1-e^{-\Delta t/\tau} }\right)g\left(t\right),
\end{equation}
where $\Delta t$ is the temporal scale, $g$ is the local equilibrium distribution function. In this work, the temporal integral solution is used as a heuristic solution, and we submit it into the original Boltzmann equation Eq.\ref{eq:s1b1}, after simplification with $\frac{Dg}{Dt}=B\left(g,g\right)$, the resultant equation becomes:
\begin{equation}
\frac{Df}{Dt}=e^{-\Delta t/\tau}B\left({ f,f }\right)+\left({ 1-e^{-\Delta t/\tau} }\right)\left[{ B\left({ f,g }\right)+B\left({ g,f }\right) }\right],
\end{equation}
where $\left[{ B\left({ f,g }\right)+B\left({ g,f }\right) }\right]$ is the collision term of the linearized Boltzmann equation (LBE), and is denoted by $L\left(f\right)$ in this work. Thus, the resultant equation can be written as:
\begin{equation}\label{eq:mbe}
\frac{Df}{Dt}=e^{-\Delta t/\tau}B\left({ f,f }\right)+\left({ 1-e^{-\Delta t/\tau} }\right)L\left({ f }\right).
\end{equation}
The assumption of LBE is the distribution function not far from non-equilibrium (near continuum), and $L\left(f\right)$ describes the molecular collisions when $\Delta t\gg\tau$ in continuum flow regime. In this work, the Eq.\ref{eq:mbe} is called the multi-scale Boltzmann equation (MBE). Examine this equation, some useful property can be found:
\begin{description}
    \item[Property (1)] The observation scale is introduced to the master equation.
    \item[Property (2)] The multi-scale collision term is the weighted average of the non-equilibrium BE one and the near-equilibrium LBE one, which are at the two limits of the scale, and describing the rarefied and continuum mechanism, respectively.
    \item[Property (3)] Their weights are determined by the ratio of the two temporal scales, viz. the observation time and the mean collision time.
\end{description}
It is obvious that the MBE equation has the correct Chapman-Enskog (C-E) expansion (asymptotic preserving property), since both BE and LBE have this property and MBE is their linear combination.

The interpretation of MBE collision term in this paper is as follows. One is that the molecules have weighted collision term in the whole observation time, the other is that the molecules follow the BE collision term during $e^{-\Delta t/\tau}\Delta t$ and follow LBE collision term during $\left({ 1-e^{-\Delta t/\tau} }\right)\Delta t$. In this paper, the later one is chosen in constructing a clear numerical algorithm.

Since the first (order) term of $L\left(f\right)$ is just the Boltzmann-BGK collision term, therefore $L\left(f\right)$ can be replace by relaxation collision term $\frac{g-f}{\tau}$ in an algorithm for simplicity. For solving the BE equation during $e^{-\Delta t/\tau}\Delta t$, the DSMC method can be used. While in this paper, the Quantified Model Competition (QMC) mechanism\cite{suwp} is used for solving the relaxation model during $\left({ 1-e^{-\Delta t/\tau} }\right)\Delta t$, where molecules can randomly classified as the free one and the colliding one, with the probabilities $e^{-\Delta t/\tau}$ and $1-e^{-\Delta t/\tau}$, respectively.
\begin{equation}
f\left({ t+\delta t }\right)=e^{-\delta t/\tau}f\left(t\right)+\left({ 1-e^{-\delta t/\tau} }\right)g_{CE}.
\end{equation}
For colliding molecules, since they follow the second order CE distribution, they can be represented by the N-S equation and drop their individual information of velocity and coordinates. Therefore, the algorithm for MBE can be rearranged as two steps:
\begin{description}
    \item[Step (1)] Molecules are classified as candidate free and colliding molecules during $e^{-\Delta t/\tau}\Delta t$. The colliding molecules are deleted and merged into the macroscopic variables for N-S equation.
    \item[Step (2)] DSMC.
\end{description}

\subsection{Multi-scale flux}
An important experience in designing multi-scale numerical methods in the FVM framework, like the UGKS and UGKWP method, is that the flux must be multi-scale. The core of multi-scale flux is utilizing the integral solution of model equation. Taking the BGK-type model as an example:
\begin{equation}\label{eq:bgk}
\frac{\partial f}{\partial t}+\bm{\xi}\cdot\frac{\partial f}{\partial \mathbf{x}}=\frac{\mathscr{G}-f}{\tau}.
\end{equation}
The integral solution of Eq.\ref{eq:bgk} is:
\begin{equation}\label{eq:integral}
f\left( {\mathbf{0},t} \right) = \frac{1}{\tau}\int^t_0 \mathscr{G}\left[{ -\bm{\xi}\left({ t-\tilde{t} }\right),\tilde{t} }\right]e^{\frac{\tilde{t}-t}{\tau}} d\tilde{t} + e^{-t/\tau}f\left( {-\bm{\xi}t,0} \right),
\end{equation}
and its flux can be calculated as: ($\mathbf{n}$ denotes the unit normal vector of interface, and $S$ is the area of interface. $\Delta t$ is the length of time step.)
\begin{equation}\label{eq:flux}
\begin{aligned}
&\mathbf{F} = \frac{1}{\Delta t}\int_0^{\Delta t} \int_{R^D} \int_{R^3} \rho\bm{\varphi} f\left( {\mathbf{0},t} \right)\bm{\xi}\cdot\mathbf{n}S d\bm{\xi} d\bm{\eta} dt\\
= & \frac{1}{\Delta t}\int_0^{\Delta t} \int_{R^D} \int_{R^3} \rho\bm{\varphi} \left\{ {\frac{1}{\tau}\int^t_0 \mathscr{G}\left[{ -\bm{\xi}\left({ t-\tilde{t} }\right),\tilde{t} }\right]e^{\frac{\tilde{t}-t}{\tau}} d\tilde{t}} + e^{-t/\tau}f\left( {-\bm{\xi}t,0} \right) \right\} \bm{\xi}\cdot\mathbf{n}S d\bm{\xi} d\bm{\eta} dt.
\end{aligned}
\end{equation}
Substitute second-order Taylor expansion of $\mathscr{G}\left[{ -\bm{\xi}\left({ t-\tilde{t} }\right),\tilde{t} }\right]$,
\begin{equation}\label{eq:taylor}
\mathscr{G}\left({ \mathbf{x},t }\right)=\mathscr{G}|_0+\frac{\partial \mathscr{G}}{\partial \mathbf{x}}|_0\cdot\mathbf{x}+\frac{\partial \mathscr{G}}{\partial t}|_0t,
\end{equation}
into Eq.\ref{eq:integral}, it can be derived into:
\begin{equation}\label{eq:expansion}
f\left( {\mathbf{0},t} \right) = \frac{1}{\tau}\int^t_0\left[{ \mathscr{G}|_0 + \frac{\partial \mathscr{G}}{\partial \mathbf{x}}|_0\cdot\left({ -\bm{\xi} }\right)\left({ t-\tilde{t} }\right)+\frac{\partial \mathscr{G}}{\partial t}|_0\tilde{t} }\right]e^{\frac{\tilde{t}-t}{\tau}} d\tilde{t} + e^{-t/\tau}f\left( {-\bm{\xi}t,0} \right).
\end{equation}
Substituting Eq.\ref{eq:expansion} into Eq.\ref{eq:flux}, it is derived as:
\begin{equation}\label{eq:core}
\begin{aligned}
&\mathbf{F} = \frac{1}{\Delta t}\int_{R^D} \int_{R^3} \rho\bm{\varphi} \left( {\delta_1\mathscr{G}|_0 + \delta_2\frac{\partial \mathscr{G}}{\partial \mathbf{x}}|_0\cdot\bm{\xi} + \delta_3\frac{\partial \mathscr{G}}{\partial t}|_0} \right) \bm{\xi}\cdot\mathbf{n}S d\bm{\xi} d\bm{\eta}\\
+ & \frac{1}{\Delta t}\int_0^{\Delta t} \int_{R^D} \int_{R^3} \rho\bm{\varphi} e^{-t/\tau}f\left( {-\bm{\xi}t,0} \right) \bm{\xi}\cdot\mathbf{n}S d\bm{\xi} d\bm{\eta} dt,
\end{aligned}
\end{equation}
where,
\begin{equation}
\begin{aligned}
\delta_1 &= \Delta t-\tau\left( {1-e^{-\Delta t/\tau}} \right),\\
\delta_2 &= 2\tau^2\left( {1-e^{-\Delta t/\tau}} \right)-\tau\Delta t-\tau\Delta te^{-\Delta t/\tau},\\
\delta_3 &= \frac{\Delta t^2}{2}-\tau\Delta t+\tau^2\left( {1-e^{-\Delta t/\tau}} \right).
\end{aligned}
\end{equation}
Eq.\ref{eq:core} is the core of success of UGKS and UGKWP method. The former term denotes the macroscopic affect on the flux. Only macroscopic variables and their gradients are needed to calculate it. The latter term denotes the free transport part. In the UGKWP method, it is described by particles. As Ref.\cite{suwp}, the former term is further rewritten as:
\begin{equation}\label{eq:ce1}
\begin{aligned}
& \frac{1}{\Delta t} \int_{R^D} \int_{R^3} \rho\bm{\varphi} \left( {\delta_1\mathscr{G}|_0 + \delta_2\frac{\partial \mathscr{G}}{\partial \mathbf{x}}|_0\cdot\bm{\xi} + \delta_3\frac{\partial \mathscr{G}}{\partial t}|_0} \right) \bm{\xi}\cdot\mathbf{n}S d\bm{\xi} d\bm{\eta}\\
= & \frac{1}{\Delta t} \int_{R^D} \int_{R^3} \rho\bm{\varphi} \left\{ {\delta_1 \left[ {\mathscr{G}|_0 - \varepsilon\tau\left({ \frac{\partial \mathscr{G}}{\partial \mathbf{x}}|_0\cdot\bm{\xi} + \frac{\partial \mathscr{G}}{\partial t}|_0} \right)} \right] + \left[ { \frac{\Delta t^2}{2}-\tau^2+\left({\tau^2+\tau\Delta t}\right)e^{-\Delta t/\tau} } \right]\frac{\partial \mathscr{G}}{\partial t}|_0 }\right\} \bm{\xi}\cdot\mathbf{n}S d\bm{\xi} d\bm{\eta}.
\end{aligned}
\end{equation}
If the high order temporal term $\left[ { \frac{\Delta t^2}{2}-\tau^2+\left({\tau^2+\tau\Delta t}\right)e^{-\Delta t/\tau} } \right]\frac{\partial \mathscr{G}}{\partial t}|_0$ is omitted, the formula becomes,
\begin{equation}\label{eq:ce2}
\frac{1}{\Delta t} \int_{R^D} \int_{R^3} \rho\bm{\varphi} \left\{ {\delta_1 \left[ {\mathscr{G}|_0 - \varepsilon\tau\left({ \frac{\partial \mathscr{G}}{\partial \mathbf{x}}|_0\cdot\bm{\xi} + \frac{\partial \mathscr{G}}{\partial t}|_0 }\right)} \right] } \right\} \bm{\xi}\cdot\mathbf{n}S d\bm{\xi} d\bm{\eta}.
\end{equation}
It is found that the distribution function here is in line with the Chapmann-Enskog expansion, which is consistent with the N-S equations. In the view of BGK model or Shakhov model, the complicated integration in Eq.\ref{eq:core} can be replaced by simple N-S solver as Ref.\cite{suwp}. Furthermore, extending the flow to multi-component flow or even thermochemical non-equilibrium flow, this term can also be calculated by corresponding N-S solver, while the BGK-type model equation for such kind of complicated flow is unavailable at this moment. The parameter,
\begin{equation}\label{eq:vp}
\varepsilon=\frac{-2\left( {1-e^{-\Delta t/\tau}} \right)+\frac{\Delta t}{\tau}\left({1+e^{-\Delta t/\tau}}\right)}{\frac{\Delta t}{\tau}-\left( {1-e^{-\Delta t/\tau}} \right)},
\end{equation}
is named as non-equilibrium correction factor for N-S solver's viscosity. When $\Delta t \gg \tau$, $\varepsilon \rightarrow 1$. As a consequence, in the process of calculating multi-scale flux, the multi-scale model is not relied on. Only N-S solver for calculating the former term in Eq.\ref{eq:core} is needed (as Eq.\ref{eq:ce2}), and free-transport particles for calculating the latter term in Eq.\ref{eq:core} is needed.

\begin{figure}
	\centering
	\subfigure{
			\includegraphics[width=0.6 \textwidth]{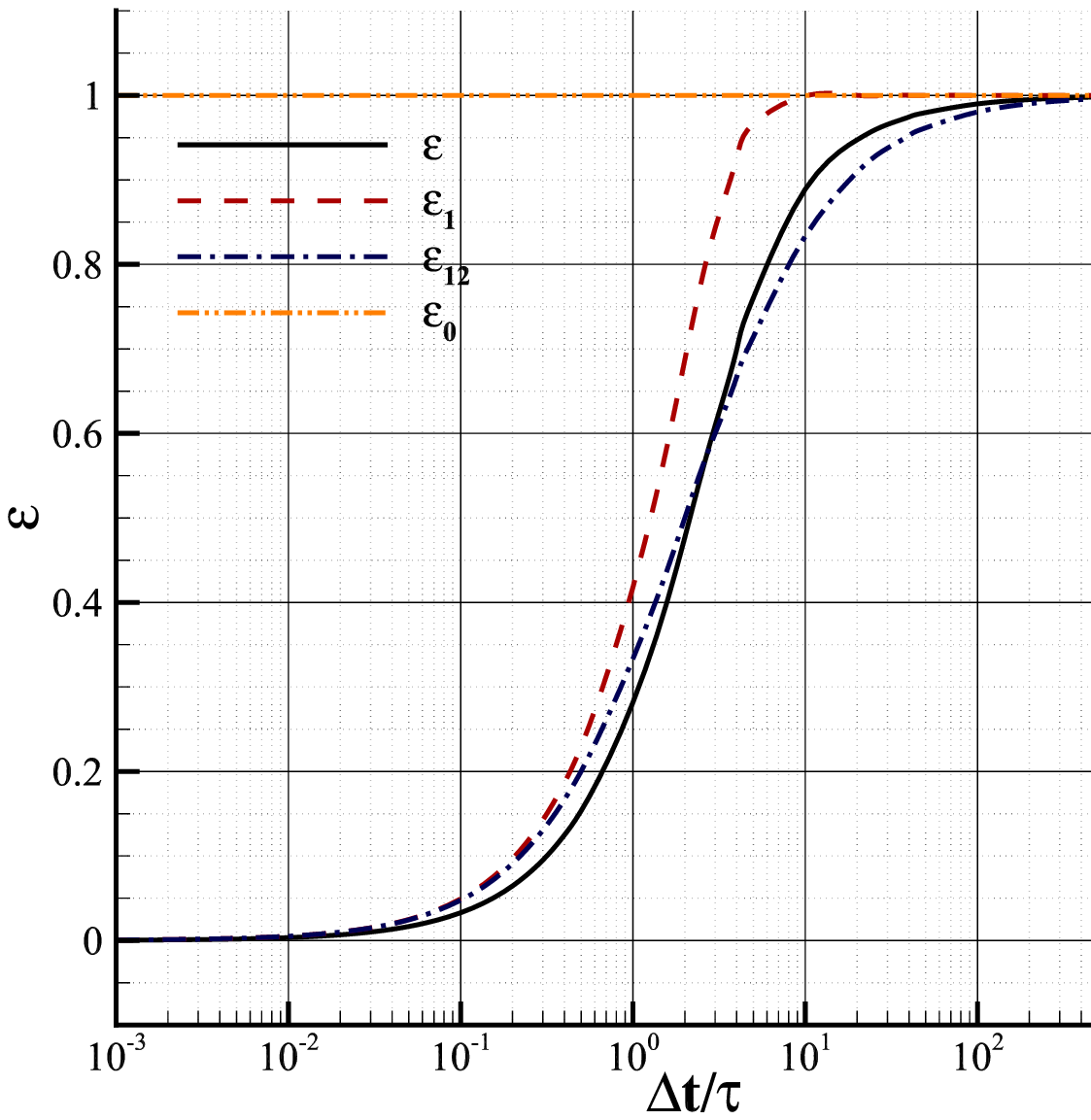}
		}
	\caption{\label{figa} Profile of $\varepsilon$.}
\end{figure}

In the DUWP method, the implicit differential method is used instead of integral formula with $exp()$ function. To match this, as Ref.\cite{suwp}, the time integration in Eq.\ref{eq:flux} is not used. Instead, the solution at $t=\Delta t$ is used, and it is derived that:
\begin{equation}
\begin{aligned}
&\mathbf{F} = \int_{R^D} \int_{R^3} \rho\bm{\varphi} f\left( {\mathbf{0},\Delta t} \right)\bm{\xi}\cdot\mathbf{n}S d\bm{\xi} d\bm{\eta}\\
= &\int_{R^D} \int_{R^3} \rho\bm{\varphi} \left({1-e^{-\frac{\Delta t}{\tau}}}\right)\left[ {\mathscr{G}|_0 - \varepsilon_1\tau\left({\frac{\partial \mathscr{G}}{\partial \mathbf{x}}|_0\cdot\bm{\xi} + \frac{\partial \mathscr{G}}{\partial t}|_0}\right) } \right] \bm{\xi}\cdot\mathbf{n}S d\bm{\xi} d\bm{\eta},
\end{aligned}
\end{equation}
where,
\begin{equation}\label{eq:vp1}
\varepsilon_1=1-\frac{\Delta t}{\tau}\frac{e^{-\frac{\Delta t}{\tau}}}{1-e^{-\frac{\Delta t}{\tau}}}.
\end{equation}
If a second-order approximation is used to $e^{-\frac{\Delta t}{\tau}}=\frac{1}{1+\frac{\Delta t}{\tau}+\frac{\Delta t^2}{2\tau^2}}$, a more simple formula can be derived from $\varepsilon_1$: $\varepsilon_{12}=\frac{\Delta t}{2\tau+\Delta t}$. The most simple parameter is that $\varepsilon_0=1$. The profile of different kinds of $\varepsilon$ is shown in Fig.\ref{figa}.

\subsection{Algorithm of DUWP}
\begin{figure}
	\centering
	\subfigure{
			\includegraphics[width=0.75 \textwidth]{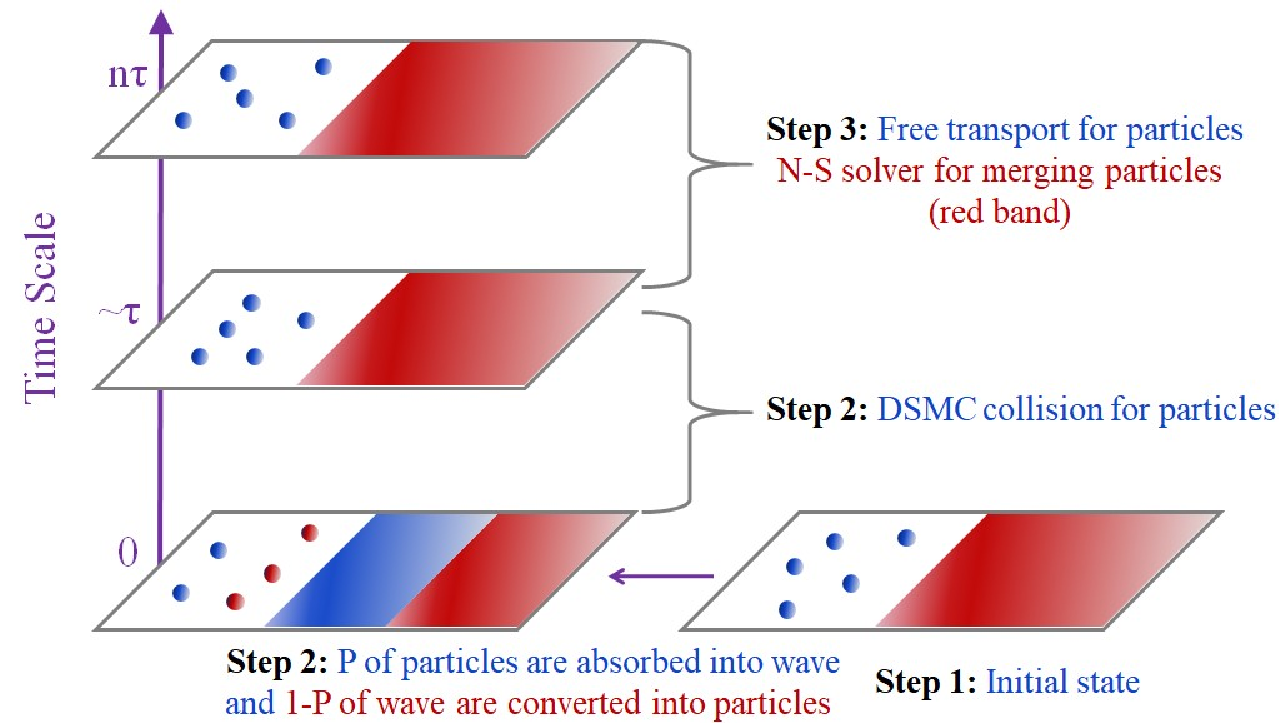}
		}
	\caption{\label{fig2} A diagrammatic drawing for DUWP.}
\end{figure}

The algorithm of DUWP is summed as follows:
\begin{description}
    \item[Step (1)] Initial state. Give an initial state with macroscopic variables and microscopic particles, and,
    \begin{equation}
    \bm{\Psi}_t = \bm{\Psi}_w + \bm{\Psi}_p,
    \end{equation}
    where subscript ``t'' denotes the total conservative macroscopic value in this cell, ``w'' denotes the conservative macroscopic value recorded by the macroscopic part (wave part), and ``p'' denotes the conservative macroscopic value counted by the stochastic particle part.
    \item[Step (2)] Collision. According to Eq.\ref{eq:p}, $P$ of particles will be absorbed into wave part (deleted) and $1-P$ of wave part will be converted into particles as:
    \begin{equation}\label{eq:maxsample}
    \begin{aligned}
    \xi_1 &= U_{t,1} + \sqrt{2RT_t}cos(2\pi rand)\sqrt{-ln(rand)},\\
    \xi_2 &= U_{t,2} + \sqrt{2RT_t}cos(2\pi rand)\sqrt{-ln(rand)},\\
    \xi_3 &= U_{t,3} + \sqrt{2RT_t}cos(2\pi rand)\sqrt{-ln(rand)},\\
    \xi_j &= \sqrt{2RT_t}cos(2\pi rand)\sqrt{-ln(rand)}.
    \end{aligned}
    \end{equation}
    Indeed, when sampling particles from the wave part, an assumption is implicated that $\mathscr{G}\approx g$. Then, the collision term of DSMC method is used to the particles part, which will be introduced in Appendix A.
    \item[Step (3)] Flux. Particles are marked and transported as:
    \begin{equation}
    \begin{aligned}
    m^{n+1} =& m^n,\\
    \bm{\xi}^{n+1} =& \bm{\xi}^n,\\
    \mathbf{x}^{n+1} =& \mathbf{x}^{n}+\bm{\xi}^n\Delta t.
    \end{aligned}
    \end{equation}
    In this equation, superscript denotes step number, $m$ and $\mathbf{x}$ denote mass and position vector of a particle, respectively. The contribution to flux of these particles is calculated as:
    \begin{equation}\label{eq:fp}
    \mathbf{F}^p_i=\sum\limits_{j\in M(i)} \left\{ {-\sum\limits_{k\in N(ij)} m_k \bm{\varphi}_k + \sum\limits_{k\in N(ji)} m_k \bm{\varphi}_k}\right\}/V_i,
    \end{equation}
    where $M(i)$ is the gather of adjacent cells of cell $i$, and $N(ij)$ is the gather of particles passing through the interface from cell $i$ to cell $j$. Then, the flux of macroscopic part is calculated by N-S solvers. After that, the macroscopic variables at next time step can be renewed as:
    \begin{equation}\label{eq:r1}
    \bm{\Psi}^{n+1}_i = \bm{\Psi}^{n}_i+\mathbf{F}^{p,n}_i - \frac{\Delta t_i}{V_i}\sum\limits_{j\in M(i)}\mathbf{F}^{w,n}_{ij}S_{ij},
    \end{equation}
    where $\mathbf{F}^{w}$ is the flux of wave part (NS solver).
    \item[Step (4)] If the simulation is not finished, go back to step $1$.
\end{description}
A diagrammatic drawing is shown as Fig.\ref{fig2}. The gray block denotes a cell. Red color and band denote wave, while blue color and spheres denote particles.

\section{Test cases}\label{sec:cases}
One-dimensional cases in monatomic argon gas are firstly tested, like shock structure and Sod shock tube. Sod shock tube is taken as a typical unsteady case and different Knudsen numbers are considered.
\subsection{Shock structure}\label{sec:ssar}
Shock structure is a typical non-equilibrium flow. When Mach number is high, its nonequilibrium feature is strong. It is a classical case to test the ability of models or numerical methods to describe non-equilibrium flows. In this case, Mach numbers are taken as $3.0$ and $8.0$, and parameters are: $\gamma=5/3$, $\rm{Pr}=2/3$. The dynamic viscosity is calculated through $\mu\sim T^{\omega}$. The size of mesh is set to be a quarter of inflow molecular mean free path $\lambda$, which is calculated through:
\begin{equation}\label{eq:lambda}
\begin{aligned}
\lambda &= \frac{1}{\beta}\frac{\mu}{p}\sqrt{\frac{RT}{2\pi}},\\
\beta  &= \frac{{5(\alpha  + 1)(\alpha  + 2)}}{{4\alpha (5 - 2\omega )(7 - 2\omega )}}.
\end{aligned}
\end{equation}

\begin{figure}
	\centering
	\subfigure[]{
			\includegraphics[width=0.45 \textwidth]{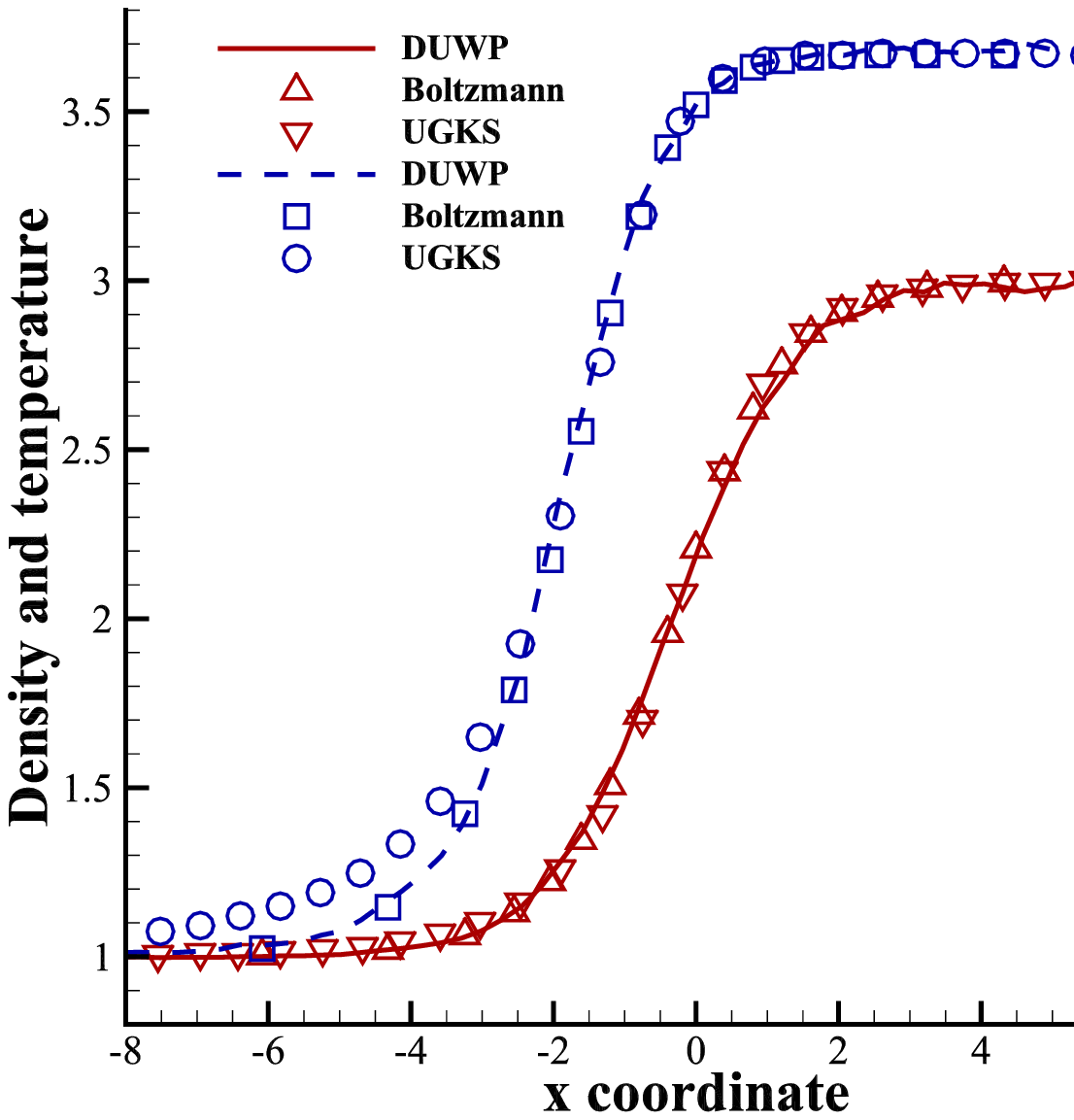}
		}
    \subfigure[]{
    		\includegraphics[width=0.45 \textwidth]{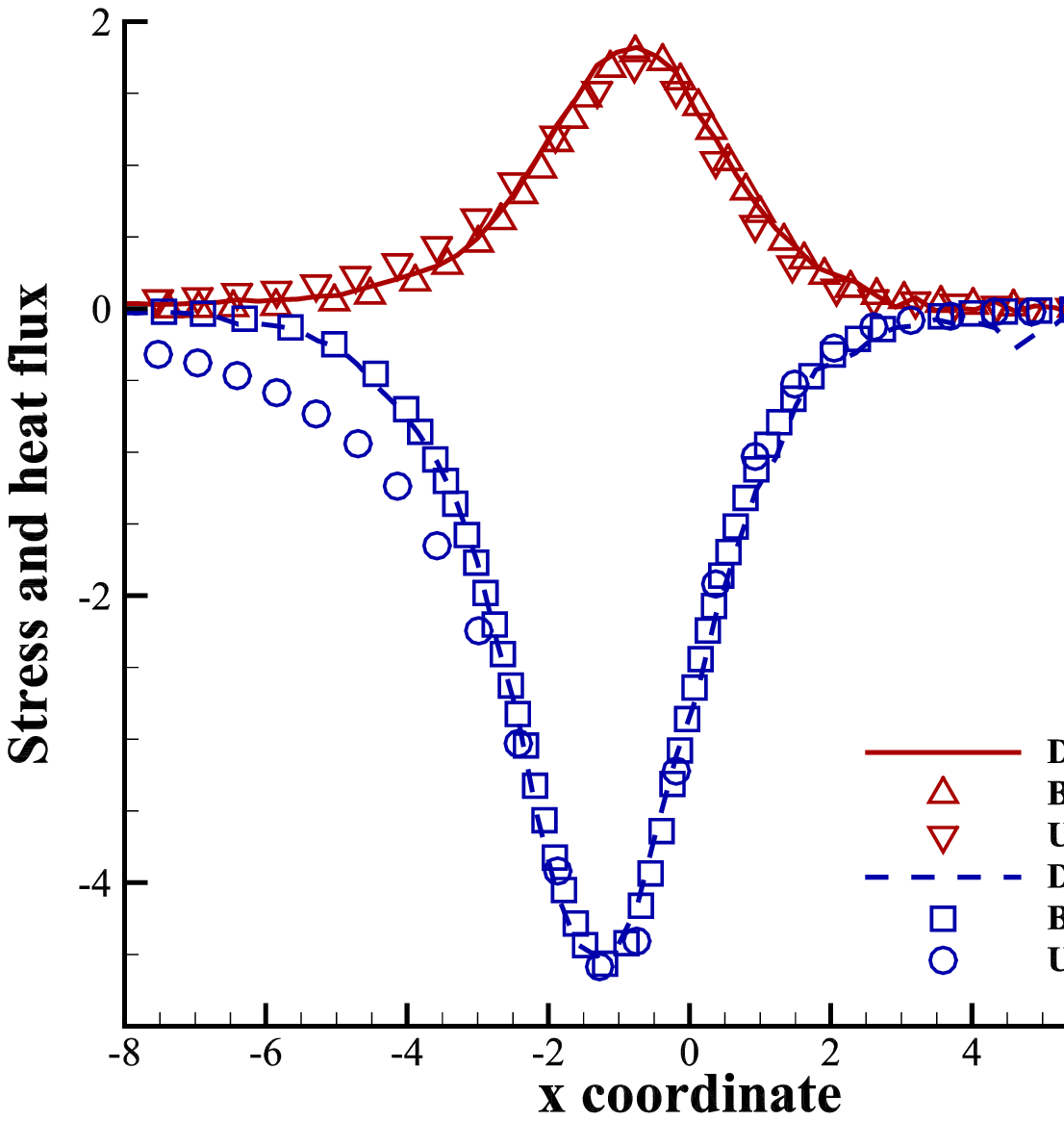}
    	}
	\caption{\label{fig3} Results of $\rm{Ma}=3.0$ argon gas shock structure: (a) Density and temperature, (b) stress and heat flux.}
\end{figure}

\begin{figure}
	\centering
	\subfigure[]{
			\includegraphics[width=0.45 \textwidth]{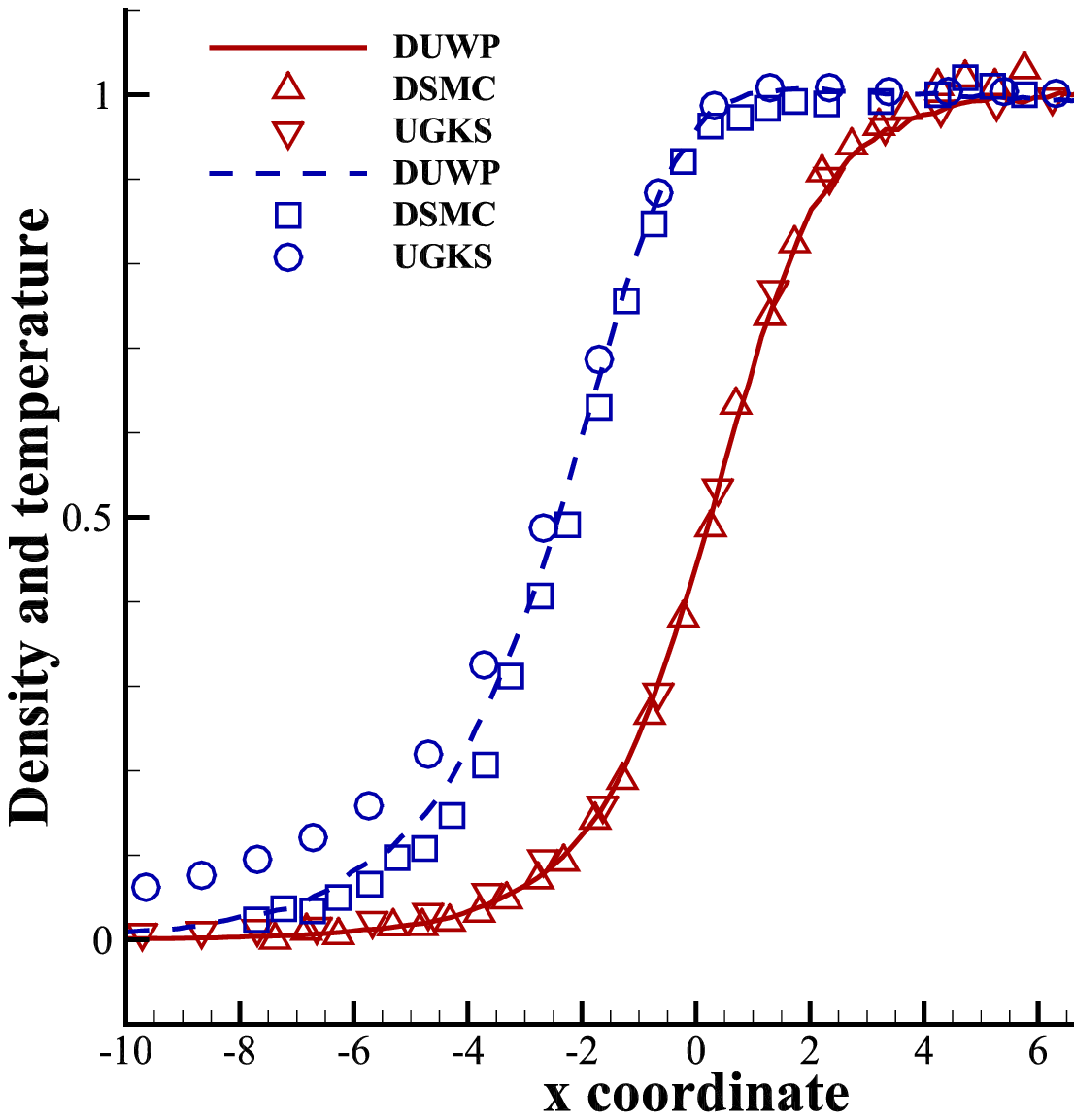}
		}
    \subfigure[]{
    		\includegraphics[width=0.45 \textwidth]{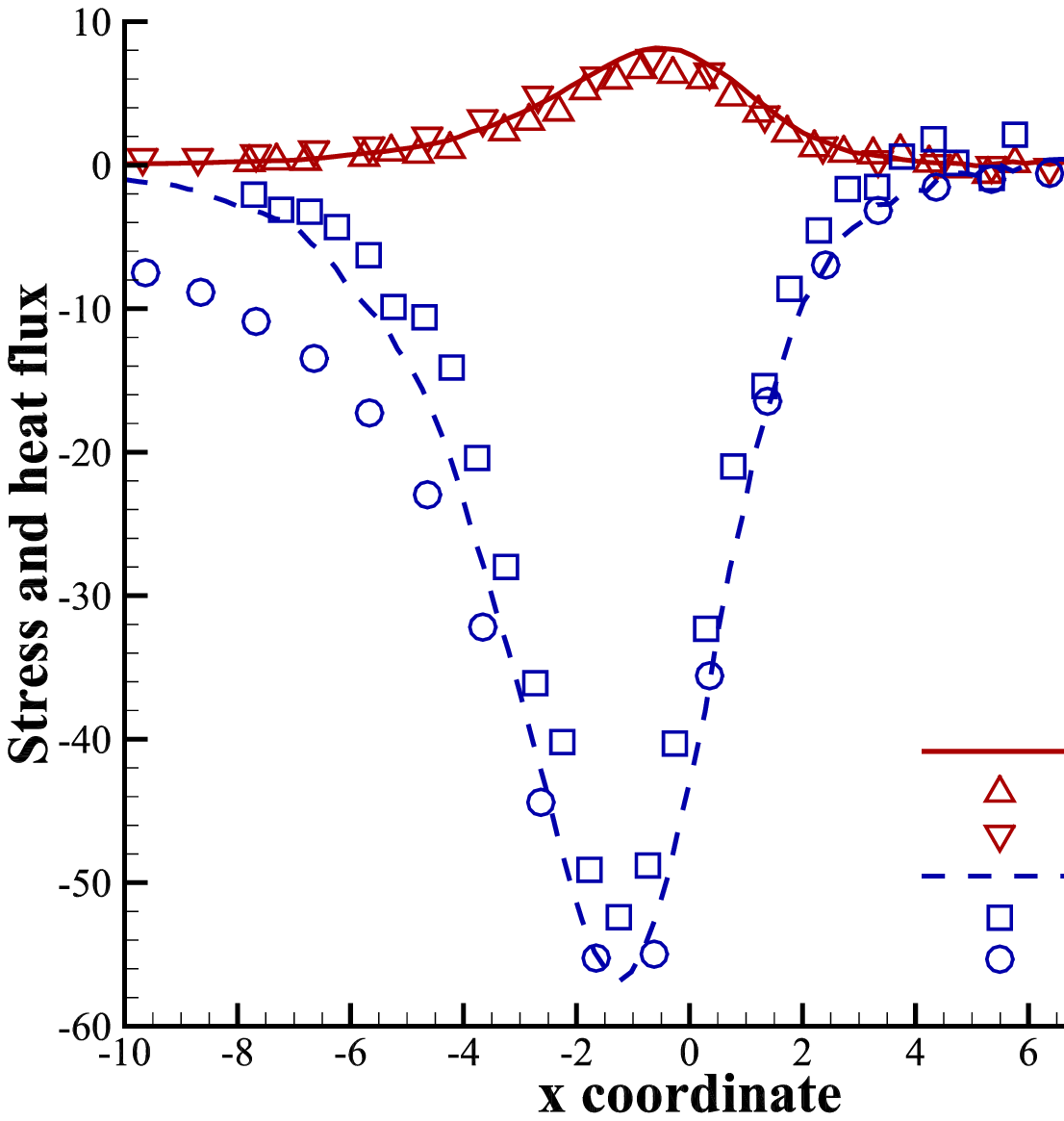}
    	}
	\caption{\label{fig4} Results of $\rm{Ma}=8.0$ argon gas shock structure: (a) Density and temperature, (b) stress and heat flux.}
\end{figure}

Other parameters are set referring to Ref.\cite{egks,dugks2015}, as follows. When Mach number is $3.0$, $\alpha=1.0$ and $\omega=0.5$. The reference data are results of Boltzmann equation\cite{boltzmannohwada} and UGKS\cite{ma3ugks}. Fig.\ref{fig3} shows the comparison results. The x coordinate is nondimensionalized through $0.5\sqrt{\pi}\lambda$. And the vertical coordinates are nondimensionalized as follows:
\begin{equation}\label{eq:ma3}
\begin{aligned}
\hat{\rho} &= \frac{\rho}{\rho_{\rm{up}}},\hat{T} &= \frac{T}{T_{\rm{up}}},\\
\hat{\Pi} &= \frac{\Pi}{p_{\rm{up}}},\hat{Q} &= \frac{Q}{p_{\rm{up}}\sqrt{2RT_{\rm{up}}}},
\end{aligned}
\end{equation}
where subscript ``up'' denotes inflow parameter. When Mach number is $8.0$, $\alpha=1.0$ and $\omega=0.68$. The reference data are results of DSMC\cite{ma8dsmc} and UGKS\cite{ma3ugks}. Fig.\ref{fig4} shows the comparison results. The x coordinate is nondimensionalized through $\beta\pi\lambda$. And the vertical coordinates are nondimensionalized as follows:
\begin{equation}\label{eq:ma8}
\begin{aligned}
\hat{\rho} &= \frac{\rho-\rho_{\rm{up}}}{\rho_{\rm{down}}-\rho_{\rm{up}}},\hat{T} &= \frac{T-T_{\rm{up}}}{T_{\rm{down}}-T_{\rm{up}}},\\
\hat{\Pi} &= \frac{\Pi}{\rho_{\rm{up}}(2RT_{\rm{up}})},\hat{Q} &= \frac{Q}{\rho_{\rm{up}}(2RT_{\rm{up}})^{1.5}},
\end{aligned}
\end{equation}
where subscript ``down'' denotes outflow parameter. The mass of a particle is $\frac{\rho_{up}\Delta x}{500}$, and $600$ steps of time average are taken. Indeed, in this case the DUWP method is equivalent to the DSMC method, because the mesh size is much smaller than $\lambda$ and $\Delta t < \tau$.

\subsection{Sod shock tube}\label{sec:sodar}
Sod shock tube is a typical non-equilibrium flow. It can test the ability of method to simulate unsteady flow, and with different inflow viscosity parameters taken, the flow regime can be set from rarefied to continuum. In this case, Knudsen numbers are taken from $1E-5$, $1E-4$, $3E-4$, $1E-3$ to $1E-2$, and parameters are: $\gamma=5/3$, $\rm{Pr}=2/3$, $\alpha=1.0$, $\omega=0.81$. The dynamic viscosity is calculated through $\mu\sim T^{\omega}$. $100$ cells are divided in the length of tube, which is the reference length $1.0$. The upstream and downstream parameters are taken as:

\begin{table}[h]
\centering
\caption{Upstream and downstream parameters of Sod shock tube}
\begin{tabular}{*{8}{|c|c|c|c|c|c|c|c}}
\hline
    $\rho_{up}$ &$u_{up}$ &$T_{up}$  &$P_{up}$  &$\rho_{down}$ &$u_{down}$ &$T_{down}$  &$P_{down}$  \\ \hline
	$1.0$       &$0.0$    &$1.0$     &$1.0$     &$0.125$       &$0.0$      &$0.8$       &$0.1$    \\
\hline
\end{tabular}
\end{table}

\begin{figure}
	\centering
	\subfigure[]{
			\includegraphics[width=0.3 \textwidth]{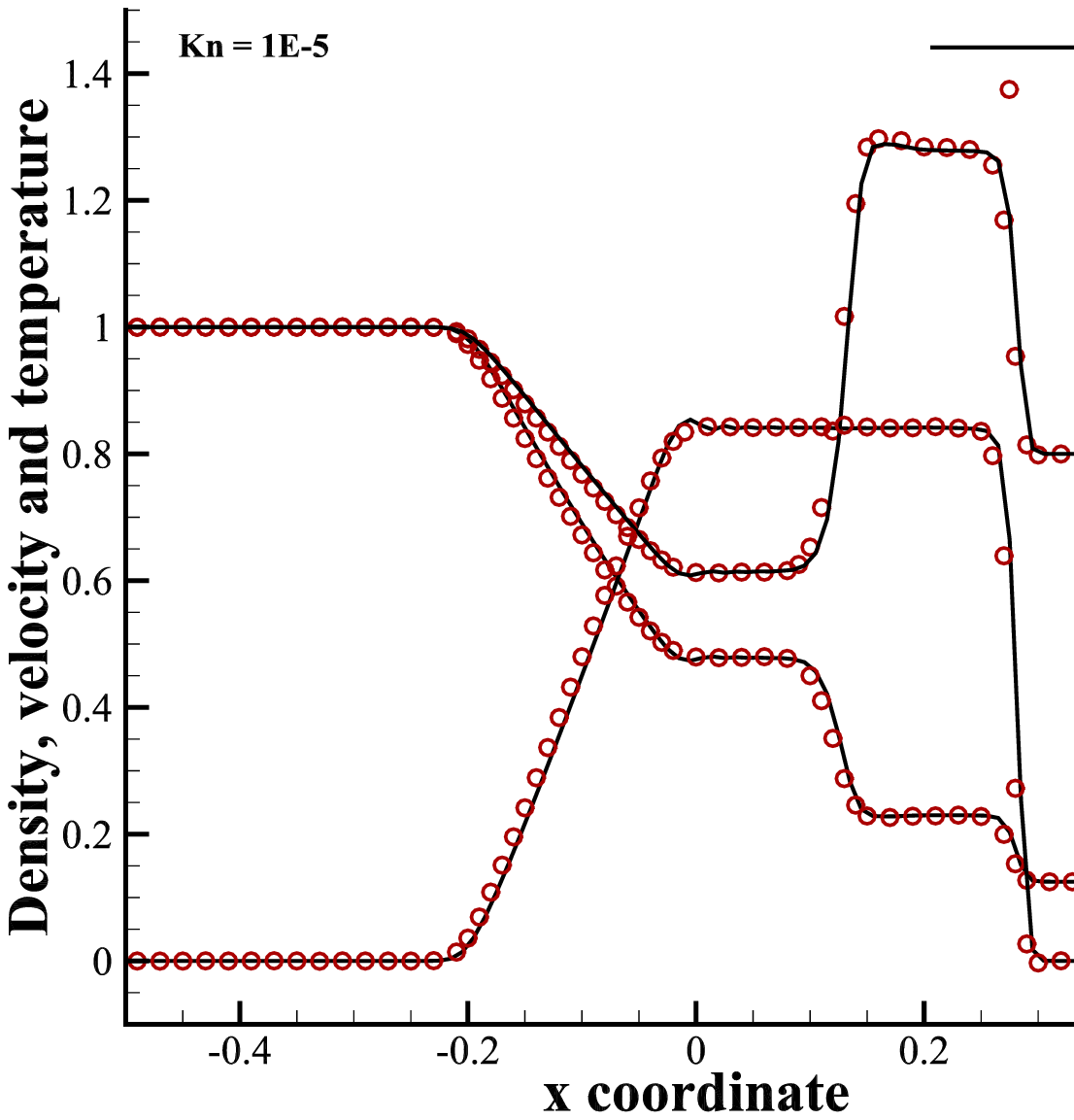}
		}
    \subfigure[]{
    		\includegraphics[width=0.3 \textwidth]{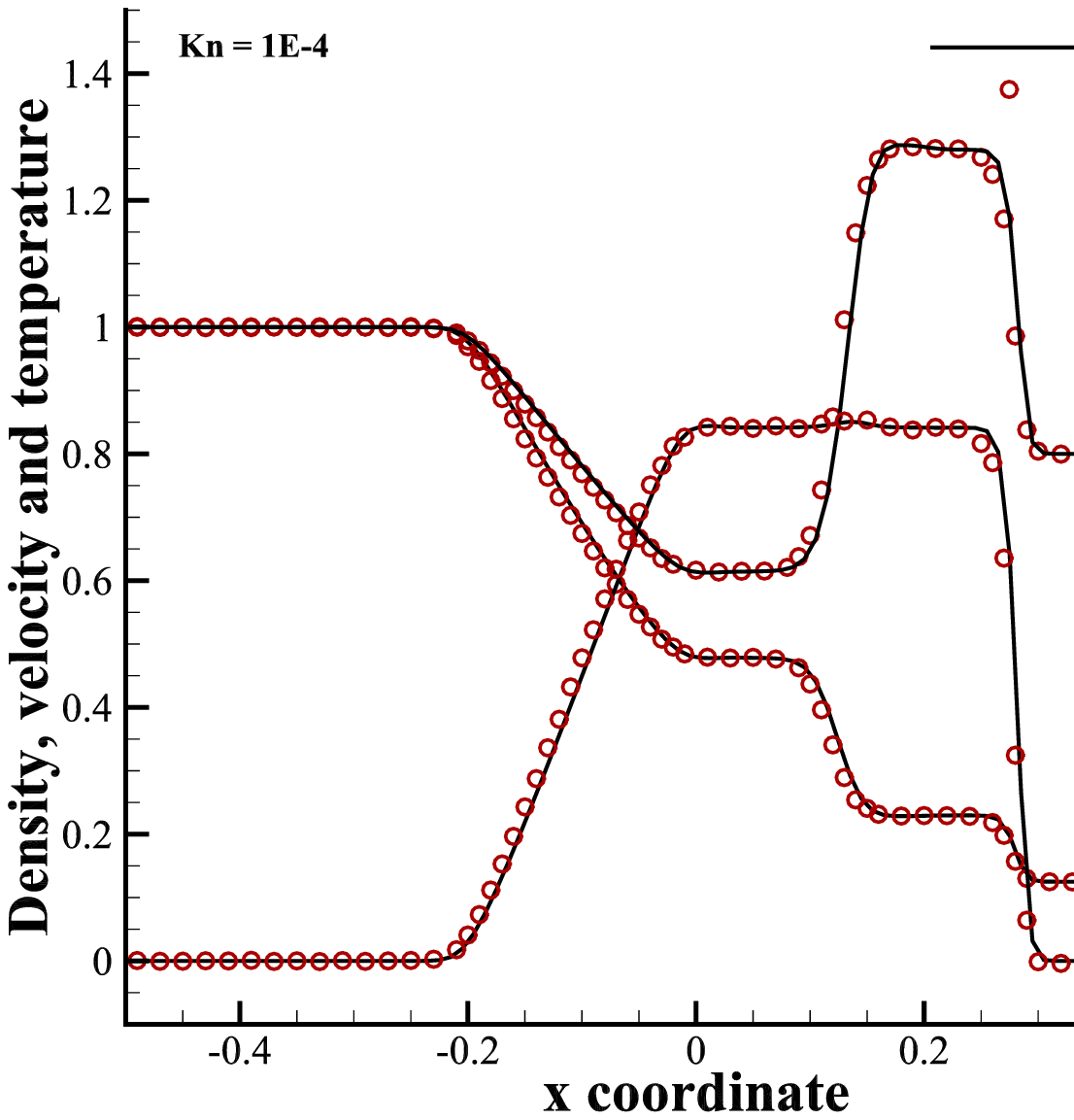}
    	}
    \subfigure[]{
    		\includegraphics[width=0.3 \textwidth]{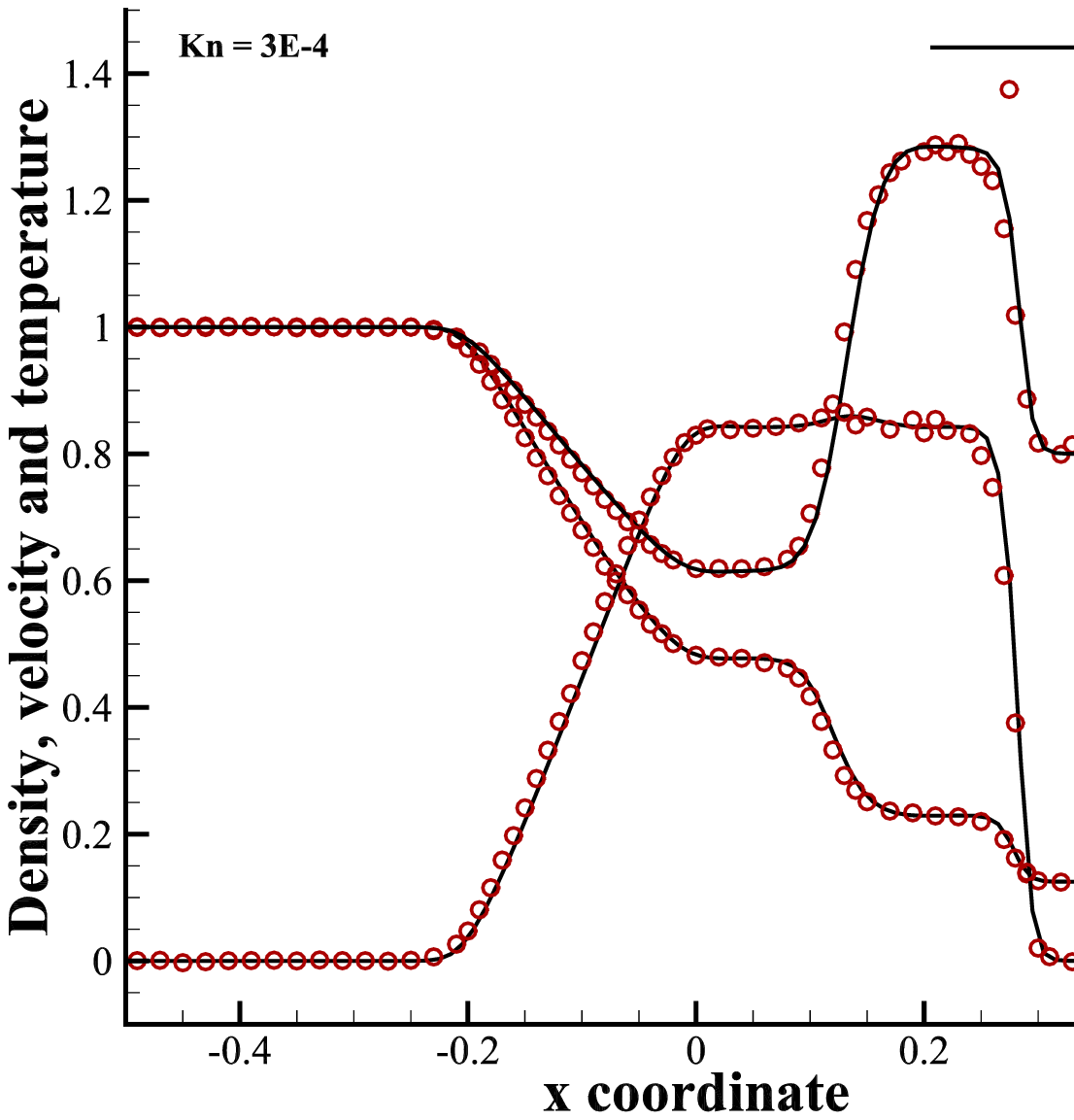}
    	}
    \subfigure[]{
    		\includegraphics[width=0.3 \textwidth]{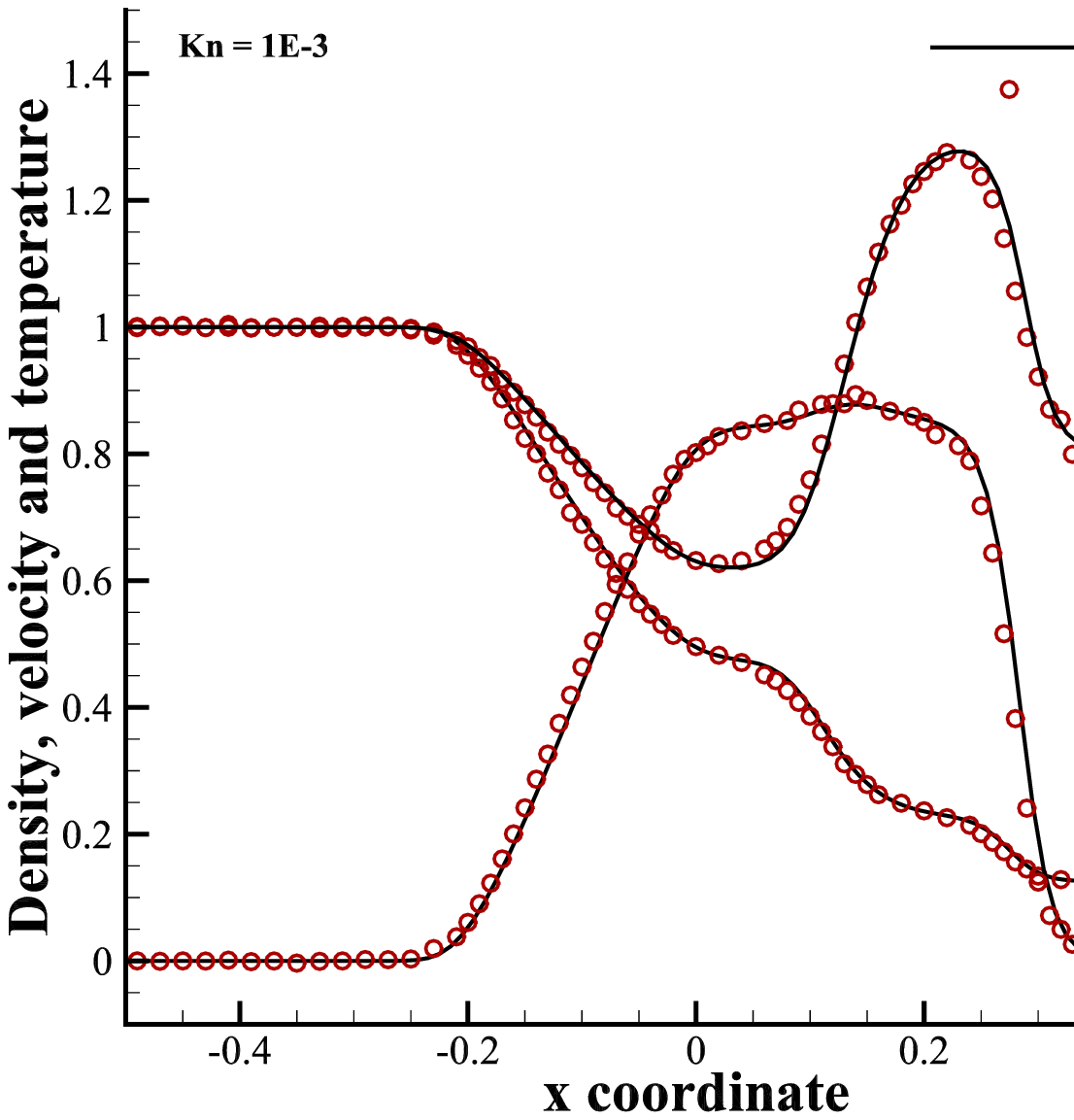}
    	}
    \subfigure[]{
    		\includegraphics[width=0.3 \textwidth]{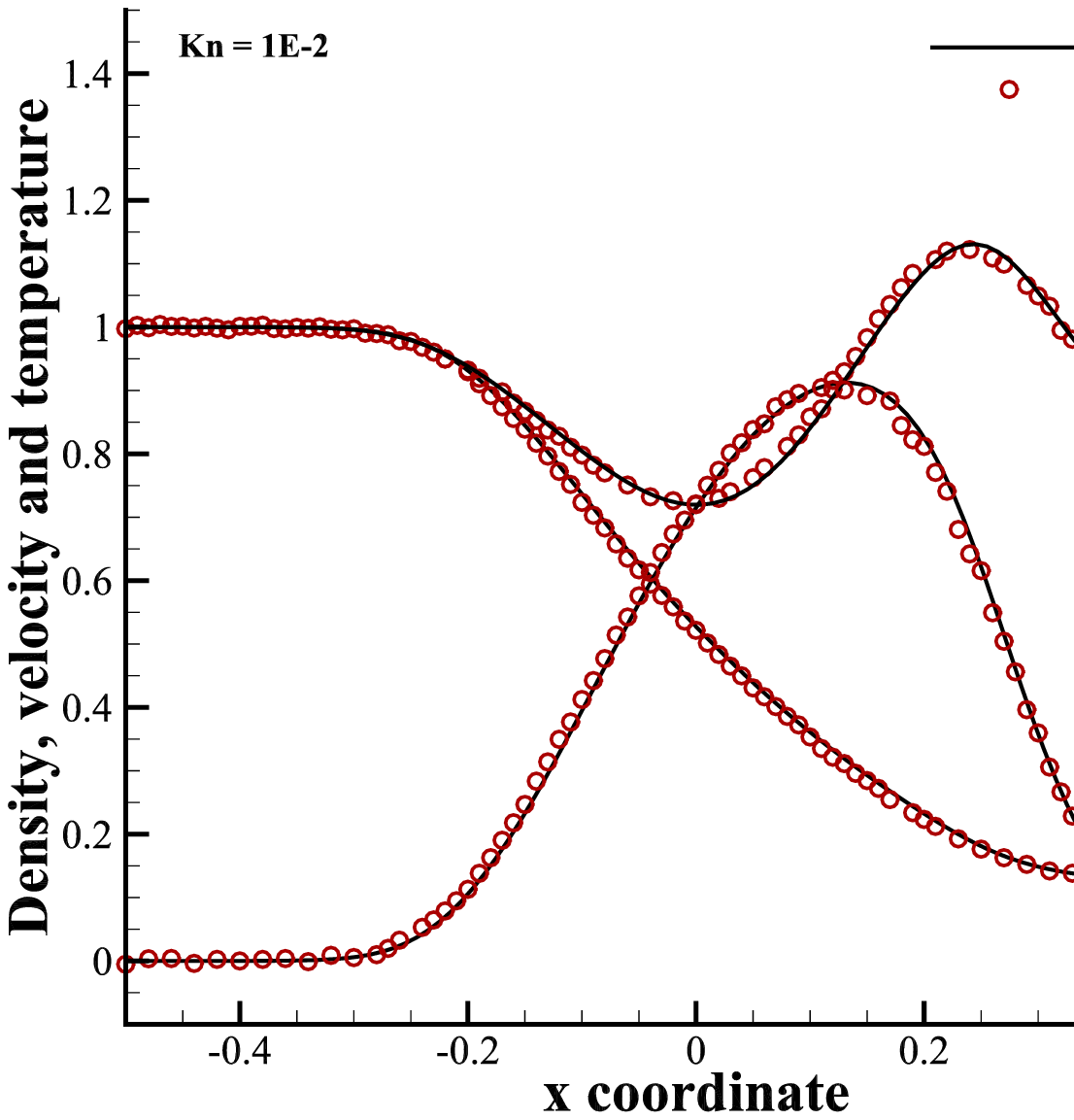}
    	}
    \subfigure[]{\label{fig5f}
    		\includegraphics[width=0.3 \textwidth]{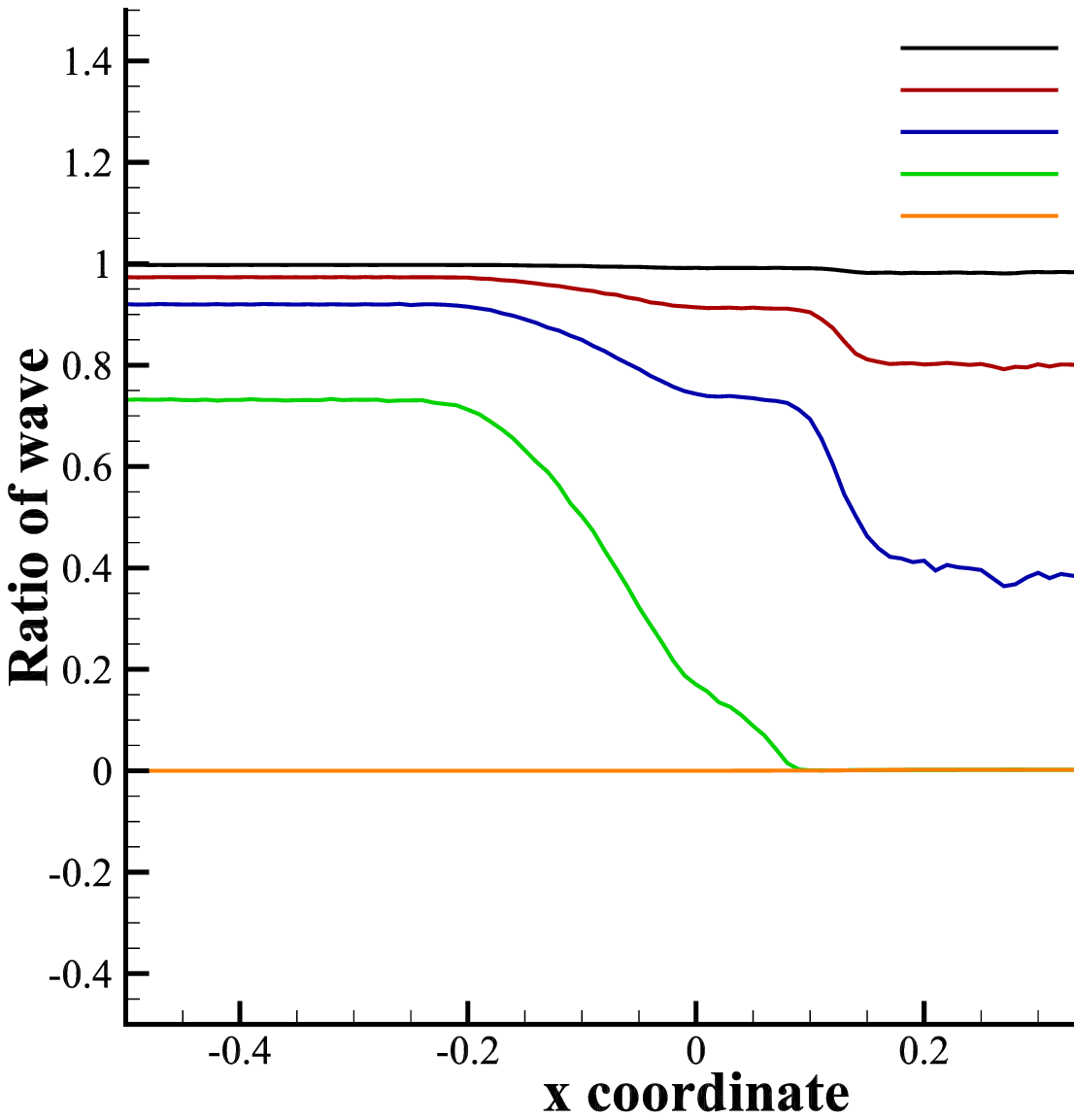}
    	}
	\caption{\label{fig5} Results of Sod shock tube: (a) $Kn=1E-5$ , (b) $Kn=1E-4$ , (c) $Kn=3E-4$ , (d) $Kn=1E-3$ , (e) $Kn=1E-2$ , (f) ratio of wave.}
\end{figure}

The mass of a particle is $\frac{\rho_{up}\Delta x}{2000}$, and $50$ steps of ensemble average are taken. Results at $t=0.15$ are shown in Fig.\ref{fig5}. Fig.\ref{fig5f} shows the ratio of wave at different Knudsen numbers. At $Kn=1E-5$, it is almost pure NS solver, and at $Kn=1E-2$ it is pure DSMC solver. Reference results are calculated by the UGKS method, and results of DUWP are in line well at each state.

\section{Conclusions}\label{sec:conclusion}
The proposed DUWP method works well in one-dimensional cases. The method will be extended to multi-dimension next. The essential point is how to deal with the great change of mesh size in the whole domain. And it is desirable to verify the accuracy of DUWP to predict stress and heat flux at the wall.

\section*{Appendix A: DSMC collision term and its nondimensionalization}
As to the collision term of DSMC, the no-time counter (NTC) method is used. For each time step, we need to sample $\frac{1}{2}N_p(N_p-1)F_N(\sigma_T\xi_r)_{max}\Delta t/V_c$ couples and decide whether they collide. As to each couple sampled, the probability of collision is $\frac{\sigma_T\xi_r}{(\sigma_T\xi_r)_{max}}$. Among the parameters, $N_p$ is the number of numerical particles; $F_N$ means that each numerical particle denotes $F_N$ real particles; $V_c$ is the volume of the cell; $\sigma_T=\pi d^2$ is the total collision cross-section; $\bm{\xi}_r=\bm{\xi}_a-\bm{\xi}_b$ is the relative velocity between the two numerical particles $a$ and $b$; $\bm{\xi}_{r,max}$ is always estimated. $d$ for variable hard sphere (VHS) model is calculated as follows, and for hard sphere (HS) model, $d$ doesn't change:
\begin{equation}
\frac{d}{d_{ref}}=\left({ \frac{\xi_r}{\xi_{r,ref}} }\right)^{-\upsilon},
\end{equation}
where $d_{ref}$ is the particle reference diameter, $\xi_{r,ref}=\sqrt{2kT_{ref}/m_r}/\Gamma(2.5-\omega)^{0.5/\upsilon}$ ($k=mR$ is the Boltzmann const,$m_r=\frac{m_am_b}{m_a+m_b}=0.5m$ is the reduced mass,$\Gamma()$ is the Gamma function), $\upsilon=\frac{2}{\eta-1}$ is the exponent of intermolecular law, and $\omega=\frac{1}{2}\frac{\eta+3}{\eta-1}$ is the viscosity index. The collision process of two particles is as follows:(superscript ``*'' denotes parameters after collision)

\begin{equation}
\begin{aligned}
&cos\theta=2rand-1, \phi=2\pi rand,\\
&\bm{\xi}^*_r=(\xi_rsin\theta cos\phi, \xi_rsin\theta sin\phi, \xi_rcos\theta),\\
&\bm{\xi}_1^*=\bm{\xi}_m+\frac{m_b}{m_a+m_b}\bm{\xi}_r^*, \bm{\xi}_2^*=\bm{\xi}_m-\frac{m_a}{m_a+m_b}\bm{\xi}_r^*,\\
&\bm{\xi}_m=\frac{m_a\bm{\xi}_a+m_b\bm{\xi}_b}{m_a+m_b}.
\end{aligned}
\end{equation}
More introductions about the NTC, collision model and corresponding parameters can be referred to monographs for DSMC\cite{dsmcre}.

A convenient nondimensionalization method is suggested as follows. Referring to the definition of mean free path:
\begin{equation}\label{eq:lambda}
\lambda=\frac{1}{\sqrt{2}\pi d^2n},
\end{equation}
where $n=F_NN/V_c$ means the number of real physical particles per volume, it is convenient to take the inflow parameters as reference:
\begin{equation}
\lambda_{ref}=\frac{1}{\sqrt{2}\pi d_{ref}^2F_{N,ref}N_{ref}/V_c}\Rightarrow d_{ref}^2=\frac{1}{\sqrt{2}\pi \lambda_{ref}F_{N,ref}N_{ref}/V_c},
\end{equation}
Then $F_{N,ref}N_{ref}/V_c$ can be counteracted in $\frac{1}{2}N(N-1)F_N(\sigma_Tc_r)_{max}\Delta t/V_c$, which is the number of sampled numerical particles per time step. So it is as follows for the VHS model:
\begin{equation}
\begin{aligned}
&\frac{1}{2}N(N-1)F_N(\sigma_Tc_r)_{max}\Delta t/V_c=\frac{1}{2\sqrt{2}\lambda_{ref}}\frac{\rho}{\rho_{ref}}(N-1)\Delta t\left({ \frac{c_{r,max}}{c_{r,ref}} }\right)^{-2\upsilon}c_{r,max},\\
&\frac{\sigma_Tc_r}{(\sigma_Tc_r)_{max}}=\frac{c_r}{(c_r)_{max}}^{1-2\upsilon}.
\end{aligned}
\end{equation}

\section*{Acknowledgements}
This work is supported by National Natural Science Foundation of China (11902264, 11902266, 12072283, 12172301) and 111 project of China (B17037). Junzhe Cao thanks Mr. Hao Jin at School of Aeronautics, Northwestern Polytechnical University for details of DSMC method.

\section*{Reference}
\bibliography{duwpref}

\newcommand{\noop}[1]{}
\begin{thebibliography}{10}
\expandafter\ifx\csname url\endcsname\relax
  \def\url#1{\texttt{#1}}\fi
\expandafter\ifx\csname urlprefix\endcsname\relax\def\urlprefix{URL }\fi
\expandafter\ifx\csname href\endcsname\relax
  \def\href#1#2{#2} \def\path#1{#1}\fi

\bibitem{pbgkre}
F.~Fei, J.~Zhang, J.~Li, Z.~Liu, A unified stochastic particle
  {Bhatnagar-Gross-Krook} method for multiscale gas flows, Journal of
  Computational Physics 400 (2020) 108972.

\bibitem{dsmcpbgkre}
F.~Fei, P.~Jenny, A hybrid particle approach based on the unified stochastic
  particle {Bhatnagar-Gross-Krook} and {DSMC} methods, Journal of Computational
  Physics 424 (2021) 109858.

\bibitem{dvmre}
L.~Mieussens, Discrete-velocity models and numerical schemes for the
  {Boltzmann-BGK} equation in plane and axisymmetric geometries, Journal of
  Computational Physics 162~(2) (2000) 429--466.

\bibitem{imexre}
W.~Boscheri, G.~Dimarco, High order central {WENO-Implicit-Explicit Runge
  Kutta} schemes for the {BGK} model on general polygonal meshes, Journal of
  Computational Physics 422 (2020) 109766.

\bibitem{ugksre}
T.~Xiao, C.~Liu, K.~Xu, Q.~Cai, A velocity-space adaptive unified gas kinetic
  scheme for continuum and rarefied flows, Journal of Computational Physics 415
  (2020) 109535.

\bibitem{dugksre}
Z.~Guo, K.~Xu, Progress of discrete unified gas-kinetic scheme for multiscale
  flows, Advances in Aerodynamics 3 (2021) 6.

\bibitem{idvmre}
L.~M. Yang, C.~Shu, W.~M. Yang, J.~Wu, An improved three dimensional implicit
  discrete velocity method on unstructured meshes for all {Knudsen} number
  flows, Journal of Computational Physics 396 (2019) 738--760.

\bibitem{gsisre}
W.~Su, L.~Zhu, P.~Wang, Y.~Zhang, L.~Wu, Can we find steady-state solutions to
  multiscale rarefied gas flows within dozens of iterations?, Journal of
  Computational Physics 407 (2020) 109245.

\bibitem{lvdsmc}
T.~M.~M. Homolle, N.~G. Hadjiconstantinou, A low-variance deviational
  simulation {Monte Carlo} for the {Boltzmann} equation, Journal of
  Computational Physics 226 (2007) 2341--2358.

\bibitem{apdsmc}
W.~Ren, H.~Liu, S.~Jin, An asymptotic-preserving {Monte Carlo} method for the
  {Boltzmann} equation, Journal of Computational Physics 276 (2014) 380--404.

\bibitem{csz1}
S.~Chen, K.~Xu, C.~Lee, Q.~Cai, A unified gas kinetic scheme with moving mesh
  and velocity space adaptation, Journal of Computational Physics 231~(20)
  (2012) 6643--6664.

\bibitem{csz2}
S.~Chen, C.~Zhang, L.~Zhu, Z.~Guo, A unified implicit scheme for kinetic model
  equations. {Part I}. {Memory} reduction technique, Science Bulletin 62 (2017)
  119--129.

\bibitem{yrf}
R.~Yuan, S.~Liu, C.~Zhong, A multi-prediction implicit scheme for steady state
  solutions of gas flow in all flow regimes, Communications in Nonlinear
  Science and Numerical Simulation 92 (2021) 105470.

\bibitem{dsmcre}
G.~A. Bird, Molecular gas dynamics and the direct simulation of gas flows,
  Clarendon Press, 1994.

\bibitem{jfm1}
R.~Prakash, L.~M. Le~Page, L.~P. McQuellin, S.~L. Gai, S.~O'Byrne, Direct
  simulation {Monte Carlo} computations and experiments on leading-edge
  separation in rarefied hypersonic flow, Journal of Fluid Mechanics 879 (2019)
  633--681.

\bibitem{jfm2}
S.~S. Sawant, V.~Theofilis, D.~A. Levin, On the synchronisation of
  three-dimensional shock layer and laminar separation bubble instabilities in
  hypersonic flow over a double wedge, Journal of Fluid Mechanics 941 (2022)
  A7.

\bibitem{sparta}
M.~Schouler, Y.~Pr{\'e}vereaud, L.~Mieussens, Survey of flight and numerical
  data of hypersonic rarefied flows encountered in earth orbit and atmospheric
  reentry, Progress in Aerospace Sciences 118 (2020) 100638.

\bibitem{dsmcfoam}
C.~White, M.~K. Borg, T.~J. Scanlon, S.~M. Longshaw, B.~John, D.~R. Emerson,
  J.~M. Reese, dsmcfoam+: {An OpenFOAM} based direct simulation {Monte Carlo}
  solver, Computer Physics Communications 224 (2018) 22--43.

\bibitem{jiri}
J.~Blazek, Computational fluid dynamics: {Principles} and applications,
  Elsevier, 2005.

\bibitem{toro}
E.~F. Toro, Riemann solvers and numerical methods for fluid dynamics, Springer,
  2009.

\bibitem{leveque}
R.~J. LeVeque, Finite volume methods for hyperbolic problems, Cambridge
  university press, 2002.

\bibitem{domaindecomp1}
T.~E. Schwartzentruber, L.~C. Scalabrin, I.~D. Boyd, A modular
  particle–continuum numerical method for hypersonic non-equilibrium gas
  flows, Journal of Computational Physics 225 (2007) 1159--1174.

\bibitem{domaindecomp2}
D.~B. Hash, H.~A. Hassan, Assessment of schemes for coupling {Monte Carlo} and
  {Navier-Stoks} solution methods, Journal of Thermophysics and Heat Transfer
  10~(2) (1996) 242--249.

\bibitem{domaindecomp3}
H.~S. Wijesinghe, R.~D. Hornung, A.~L. Garcia, N.~G. Hadjiconstantinou,
  Three-dimensional hybrid continuum-atomistic simulations for multiscale
  hydrodynamics, Journal of Fluids Engineering 126~(5) (2004) 768--777.

\bibitem{domaindecomp4}
Q.~Sun, I.~D. Boyd, G.~V. Candler, A hybrid continuum/particle approach for
  modeling subsonic, rarefied gas flows, Journal of Computational Physics 194
  (2004) 256--277.

\bibitem{domaindecomp5}
P.~Degond, G.~Dimarco, L.~Mieussens, A moving interface method for dynamic
  kinetic–fluid coupling, Journal of Computational Physics 227 (2007)
  1176--1208.

\bibitem{ugkwp1}
C.~Liu, Y.~Zhu, K.~Xu, Unified gas-kinetic wave-particle methods {I}:
  {Continuum} and rarefied gas flow, Journal of Computational Physics 401
  (2020) 108977.

\bibitem{ugkwp2}
Y.~Zhu, C.~Liu, C.~Zhong, K.~Xu, Unified gas-kinetic wave-particle methods.
  {II}. {Multiscale} simulation on unstructured mesh, Physics of Fluids 31
  (2019) 067105.

\bibitem{ugks1}
K.~Xu, J.-C. Huang, A unified gas-kinetic scheme for continuum and rarefied
  flows, Journal of Computational Physics 229~(20) (2010) 7747--7764.

\bibitem{ugkwp5}
X.~Xu, Y.~Chen, C.~Liu, Z.~Li, K.~Xu, Unified gas-kinetic wave-particle methods
  {V}: {Diatomic} molecular flow, Journal of Computational Physics 442 (2021)
  110496.

\bibitem{suwp}
S.~Liu, C.~Zhong, M.~Fang, Simplified unified wave-particle method with
  quantified model-competition mechanism for numerical calculation of
  multiscale flows, Physical Review E 102~(1) (2020) 013304.

\bibitem{suwp2}
W.~Liu, C.~Shu, C.~Zhang, Z.~Yuan, Y.~Liu, Z.~Zhang, A simple
  hydrodynamic-particle method for supersonic rarefied flows, Physics of Fluids
  34~(5) (2022) 057101.

\bibitem{gks}
K.~Xu, A gas-kinetic {BGK} scheme for the {Navier–Stokes} equations and its
  connection with artificial dissipation and {Godunov} method, Journal of
  Computational Physics 171~(20) (2001) 289--335.

\bibitem{ugkwpa}
X.~Yang, S.~Wei, K.~Xu, Unified gas-kinetic wave–particle method for
  gas–particle two-phase flow from dilute to dense solid particle limit,
  Physics of Fluids 34 (2022) 023312.

\bibitem{ugkwpb}
X.~Yang, Y.~Wei, S.~Wei, K.~Xu, Unified gas-kinetic wave-particle method for
  three-dimensional simulation of gas-particle fluidized bed, Chemical
  Engineering Journal 453 (2023) 139541.

\bibitem{gm3}
P.~Andries, K.~Aoki, B.~Perthame, A consistent {BGK}-type model for gas
  mixtures, Journal of Statistical Physics 106~(5/6) (2002) 993--1018.

\bibitem{gm4}
Y.~Zhang, L.~Zhu, R.~Wang, Z.~Guo, Discrete unified gas kinetic scheme for all
  {Knudsen} number flows. {III. Binary} gas mixtures of {Maxwell} molecules,
  Physical Review E 97~(5) (2018) 053306.

\bibitem{gm5}
M.~Groppi, S.~Monica, G.~Spiga, A kinetic ellipsoidal {BGK} model for a binary
  gas mixture, Europhysics Letters 96~(6) (2011) 64002.

\bibitem{gm6}
S.~Brull, An ellipsoidal statistical model for gas mixtures, Communications in
  Mathematical Sciences 13~(1) (2015) 1--13.

\bibitem{gm8}
B.~N. Todorova, R.~Steijl, Derivation and numerical comparison of {Shakhov} and
  ellipsoidal statistical kinetic models for a monoatomic gas mixture, European
  Journal of Mechanics - B/Fluids 76 (2019) 390--402.

\bibitem{gm9}
B.~N. Todorova, C.~White, R.~Steijl, Numerical evaluation of novel kinetic
  models for binary gas mixture flows, Physics of Fluids 32~(1) (2020) 016102.

\bibitem{gm11}
M.~Pirner, A {BGK} model for gas mixtures of polyatomic molecules allowing for
  slow and fast relaxation of the temperatures, Journal of Statistical Physics
  173~(6) (2018) 1660--1687.

\bibitem{gm12}
C.~Baranger, Y.~Dauvois, G.~Marois, J.~Math{\'e}, J.~Mathiaud, L.~Mieussens, A
  {BGK} model for high temperature rarefied gas flows, European Journal of
  Mechanics - B/Fluids 80 (2020) 1--12.

\bibitem{gm13}
M.~Bisi, M.~J. C{\'a}ceres, A {BGK} relaxation model for polyatomic gas
  mixtures, Communications in Mathematical Sciences 14~(2) (2016) 297--325.

\bibitem{gm14}
M.~Bisi, R.~Travaglini, A {BGK} model for mixtures of monoatomic and polyatomic
  gases with discrete internal energy, Physica A 547 (2020) 124441.

\bibitem{gm15}
M.~Bisi, R.~Monaco, A.~Soares, A {BGK} model for reactive mixtures of
  polyatomic gases with continuous internal energy, Journal of Physics A:
  Mathematical and Theoretica 51~(12) (2018) 125501.

\bibitem{dr}
S.~Yang, S.~Liu, C.~Zhong, J.~Cao, C.~Zhuo, A direct relaxation process for
  particle methods in gas-kinetic theory, Physics of Fluids 33 (2021) 076109.

\bibitem{xiaocong}
X.~Xu, Y.~Chen, K.~Xu, Modeling and computation for non-equilibrium gas
  dynamics: {Beyond} single relaxation time kinetic models, Physics of Fluids
  33~(1) (2021) 011703.

\bibitem{dm}
K.~Xu, Direct modeling for computational fluid dynamics: Construction and
  application of unified gas-kinetic schemes, World Scientific, 2015.

\bibitem{egks}
S.~Liu, Y.~Yang, C.~Zhong, An extended gas-kinetic scheme for shock structure
  calculations, Journal of Computational Physics 390 (2019) 1--24.

\bibitem{dugks2015}
Z.~Guo, R.~Wang, K.~Xu, Discrete unified gas kinetic scheme for all {Knudsen}
  number flows. {II}. {Thermal} compressible case, Physical Review E 91~(3)
  (2015) 033313.

\bibitem{boltzmannohwada}
T.~Ohwada, Structure of normal shock waves: {Direct} numerical analysis of the
  {Boltzmann} equation for hard‐sphere molecules, Physics of Fluids A: Fluid
  Dynamics 5~(1) (1993) 217.

\bibitem{ma3ugks}
K.~Xu, J.-C. Huang, An improved unified gas-kinetic scheme and the study of
  shock structures, IMA Journal of Applied Mathematics 76~(5) (2011) 698 --
  711.

\bibitem{ma8dsmc}
G.~A. Bird, Aspects of the structure of strong shock waves, The Physics of
  Fluids 13~(5) (1970) 1172.

\end{thebibliography}

\end{document}